# Chemical
# Science

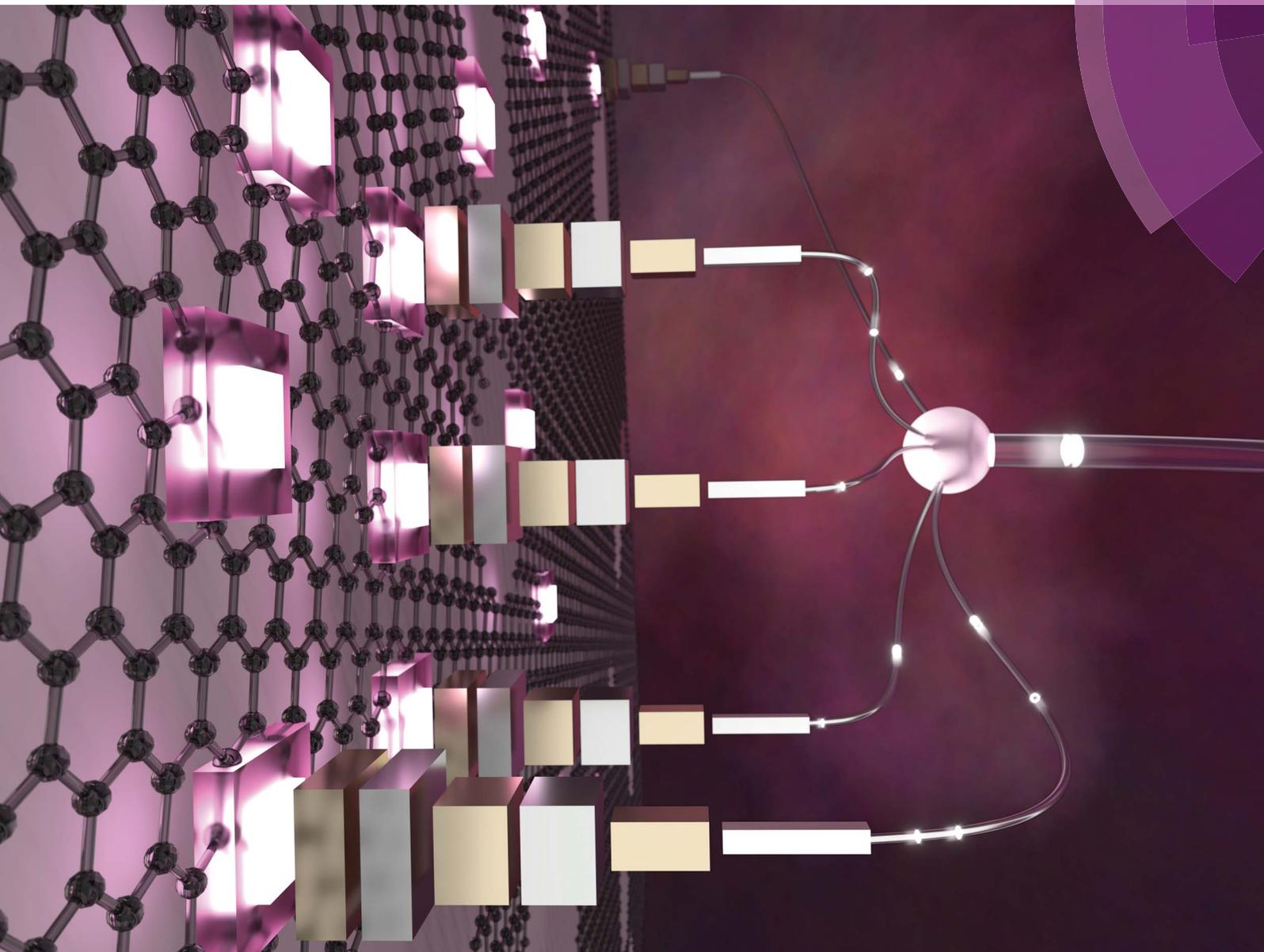

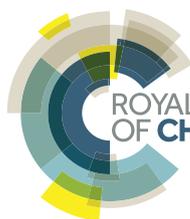

ROYAL SOCIETY
OF CHEMISTRY | Celebrating IYPT 2019

**EDGE ARTICLE**
Kyle Mills *et al.*
Extensive deep neural networks for transferring small scale
learning to large scale systems

## EDGE ARTICLE

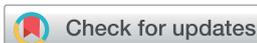 Check for updates

# Extensive deep neural networks for transferring small scale learning to large scale systems

Kyle Mills,  *[a] Kevin Ryczko, [b] Iryna Luchak,[c] Adam Domurad,[d] Chris Beeler [a] and Isaac Tamblyn [abe]

We present a physically-motivated topology of a deep neural network that can efficiently infer extensive parameters (such as energy, entropy, or number of particles) of arbitrarily large systems, doing so with $\mathcal{O}(N)$ scaling. We use a form of domain decomposition for training and inference, where each sub-domain (tile) is comprised of a non-overlapping focus region surrounded by an overlapping context region. The size of these regions is motivated by the physical interaction length scales of the problem. We demonstrate the application of EDNNs to three physical systems: the Ising model and two hexagonal/graphene-like datasets. In the latter, an EDNN was able to make total energy predictions of a 60 atoms system, with comparable accuracy to density functional theory (DFT), in 57 milliseconds. Additionally EDNNs are well suited for massively parallel evaluation, as no communication is necessary during neural network evaluation. We demonstrate that EDNNs can be used to make an energy prediction of a two-dimensional 35.2 million atom system, over 1.0 μm² of material, at an accuracy comparable to DFT, in under 25 minutes. Such a system exists on a length scale visible with optical microscopy and larger than some living organisms.

## 1 Introduction

Within the past decade, the fields of artificial intelligence, computer vision, and natural language processing have advanced at unprecedented rates. Computerized identification and classification of images, video, audio, and written text have all improved to the extent they are now part of everyday technologies. With the recent advances in hardware acceleration,[1,2] deep neural networks have been at the forefront of these developments due to their ability to perform "featureless-learning", automatically learning both the features and the mapping between raw data and quantities of interest.[3–5]

Machine learning methods are rapidly being adopted by chemists, physicists, and materials scientists, and have performed well at making predictions in the fields of dynamical mean-field theory, many-body physics,[6,7] strongly correlated materials,[8–10] phase transitions and classification,[11–15] and materials exploration and design.[16–23] Machine learning models have been shown to be of sufficient accuracy to provide fast and accurate chemical insights.[20,24–28]

[a]University of Ontario Institute of Technology, Oshawa, Ontario, Canada. E-mail: kyle.mills@uoit.net; isaac.tamblyn@nrc.ca

[b]University of Ottawa, Ottawa, Ontario, Canada

[c]University of British Columbia, Vancouver, British Columbia, Canada

[d]University of Waterloo, Waterloo, Ontario, Canada

[e]National Research Council Canada, Ottawa, Ontario, Canada

Convolutional deep neural networks have been used to predict the kinetic energy of hydrocarbons and were successful in reproducing the Kohn–Sham potential energy surfaces,[29] and have been used to classify reciprocal-space diffraction patterns for crystal lattices.[30] Deep neural networks have proven their classification power in astronomical applications[31] and particle physics applications[32–34] but have yet to be widely adopted throughout the physics community for accurate numerical predictions. There are a growing number of methods being proposed to capture relevant chemistry within representations of atomic environments[35] and a consensus is forming calling for the need to incorporate physics into network design and utilize the physics of the underlying problem to motivate the use of specific network structures and techniques.[36–38] The work of Brockherde *et al.*[39] focuses on low dimensional systems and small molecules. In addition their architecture is specialized to either work with the external potential or electron density using kernel ridge regression. These models are highly specialized and do not do any sort of energy partitioning. Their respective runtime is based on the computational complexity of kernel ridge regression for training. Depending on the dimensionality of the input, we have found that kernel ridge regression requires large amounts of RAM, whereas the required memory is less for deep neural networks (DNNs).

DNNs have the ability to replace both classical[40] and quantum mechanical operators.[41,42] In comparison to other machine learning methods, convolutional deep neural networks prevailed as the most accurate and best-scaling method for all but the most simple cases.[41,43] Deep neural networks were able to learn the mapping from spin configuration to energy for multiple cases of Ising-like classical spin models. For the case of a confined quantum particle, a deep neural network successfully learned the energy of the ground state, first excited state, and kinetic energy.[41] A similar approach was able to map the structure of two dimensional hexagonal crystal lattice energies computed within the density functional theory framework.[42] All of this was accomplished *via* the aforementioned "featureless" deep learning; the network was presented with raw spatial data, without any preliminary attempt at manual feature selection.

Traditionally with deep neural networks, the training process employed is not transferable to systems of arbitrary length scales. In practice, this means that for square, $L \times L$ lattice systems, a model trained on $4 \times 4$ configurations cannot be used on an $8 \times 8$ cell (or *vice versa*) without retraining at least part of the network. With smaller or larger inputs impossible, this size limitation is clearly a major shortcoming of deep neural network techniques. Beyond the practical limitations, from a fundamental standpoint it is unsatisfying to use a model that has no concept of extensivity.

A physical property is extensive if it can be divided among subsystems. A common example is the number of particles in a system. When a sample is divided evenly into two subsystems, the number of particles in each subsystem is halved. This is in contrast to intensive quantities such as temperature that are unchanged by subdivision or addition of subsystems.

Maintaining extensivity has not been a focus of machine learning researchers, as in traditional vision- and audio-based applications of deep learning, extensivity is not a common requirement. Most classification problems (*e.g.* identification of an animal or shape in an image) are invariant to the physical dimensions of an image (*e.g.* number of pixels that comprise a cat). Indeed, absolute scale is not normally recorded in a photograph, and therefore scale invariant models are necessary and commonplace. Furthermore, photographs, handwriting, and audio recordings too large to be processed by the deep neural network can be resized, cropped, and segmented without destroying the features necessary to make a prediction;[44] there is no absolute spatial scale upon which the label of interest depends.

In a physical measurement or simulation however, the physical scale of a pixel matters critically. Consider the case of an X-ray diffraction experiment where the interference pattern recorded at the detector depends strongly on the wavelength of scattered light and the physical length scale over which the signal is collected. It is not possible to reconstruct the signal properly unless such parameters are known and are consistent. Extensivity is critical in describing chemical systems; configuration interaction with single and double excitations (CISD) is infamous for its lack of extensivity.

In this work, we propose a general method that preserves the extensivity of physical quantities, and also accommodates arbitrary input size. We propose a new deep neural network structure, which once trained, can operate on (effectively) arbitrary-sized inputs and length scales while maintaining the physical requirement of extensivity. Unlike atom-centred approaches, we avoid the problem of energy assignment or projection onto specific atoms by forcing the neural network to automatically learn by viewing the entire structure at once.

We call our approach Extensive Deep Neural Networks (EDNNs), employing domain decomposition to solve the problem of operator evaluation across length scales. Although domain decomposition techniques have a long history in computer simulation and modelling, here we have taken a new approach and allow the model itself to identify and self-optimize the overlap of tiles at domain boundaries.

Previous work[45] has identified the necessity of extensivity.[46,47] Our method is sufficiently general to allow for applications to any system in which extensivity holds, such as the spin and atomic systems we demonstrate in this work and even the charge density (*e.g.* a scalar field representing an extensive quantity). Furthermore, our method results in a model that can be evaluated on arbitrarily large system sizes, which we demonstrate.

## 2 Extensive deep neural networks

The overall objective of an EDNN is to learn the mapping between an input structure and one or more extensive properties, $\varepsilon$ (*e.g.* total energy, entropy, magnetization, particle number, charge, *etc.*). To date, our input structures have consisted of continuous, regular, real-space grids, analogous to grayscale images.

Now, one might ask: "If the property we wish to predict is extensive, can we not just split the input into blocks and add up the individual answers?" Fig. 1 provides an example. If the goal of the neural network is to count the number of dots in the box (left), then division into non-overlapping subsystems will indeed yield the correct answer. If, however, the task is to count the number of multicolored pairs (right), the process is not so straightforward, and subdivision without accounting for the boundary yields erroneous answers.

The spatial extent over which features in the configuration influence the value of an operator is known as locality. In general, operators (such as the number operator, Hamiltonian, magnetization, *etc.*) may be described as local, semi-local, or fully non-local. In the density functional theory framework, exchange–correlation functionals are often identified in these categories. The local density approximation is considered local, generalized gradient approximations like PW91 [50] and PBE [51] considered semi-local and some hybrid functionals (*e.g.* B3LYP [52,53]) and exact exchange considered non-local.

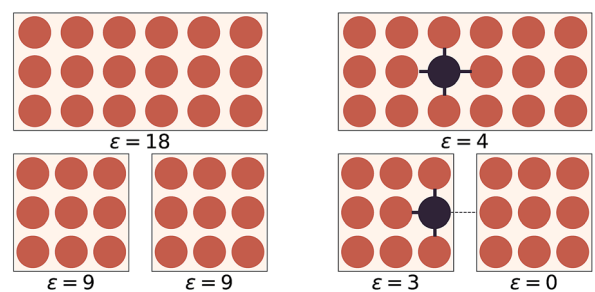

**Fig. 1** On the left, the counting operator is local and the sum of the operator applied to individual subsystems results in the same answer as the operator applied to the complete system. On the right, the semi-local nearest-neighbour operator (*i.e.* the count of the number of black-red neighbours) cannot be applied to subsystems separately and then summed.

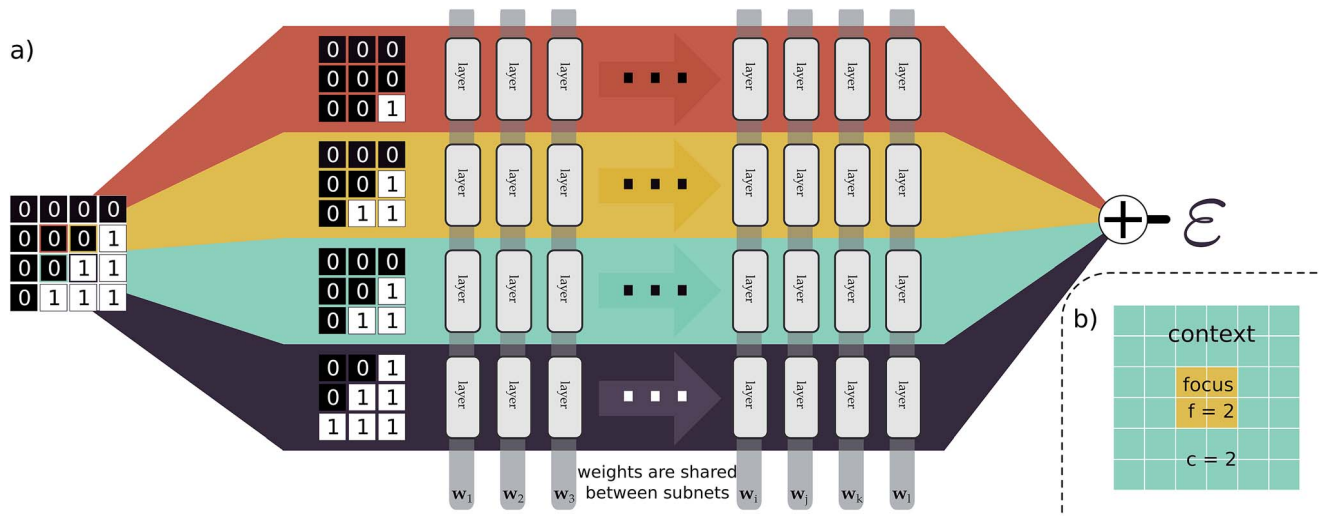

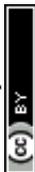

**Fig. 2** An input example is decomposed into four tiles, with each tile consisting of a focus and context region. (a) As a pedagogical example, we expand 4 adjacent tiles comprising a generic binary grid. For this case both the focus and the context are unit width, resulting in 3 × 3 tiles. The tiles are simultaneously passed through the same neural network (*i.e.* the same weights). The individual outputs are summed, producing an estimate of $\varepsilon$, an extensive quantity. When training, the cost function is assessed after this summation, forcing the weight updates to consider all input tiles simultaneously. In (b), we show an example tile with a focus of $f = 2$, and a context of $c = 2$. The optimal selection of $f$ and $c$ depend on the physical length scale of the target (learned) function.

We define $l$ to be the length scale of an operator's locality. For example, in the counting example above, the number operator is fully local ($l = 0$), as computing $\varepsilon$ requires only local knowledge. The nearest-neighbour example is non-local with $l = 1$, meaning knowledge of the surrounding region is necessary to make a prediction. The gradient operator using a second-order finite difference method is an example of a semi-local operator with $l = 2$.

For many systems, such as the Coulomb ($1/r$) interaction, there is no hard cut off, but typically one expects the importance of a feature to diminish as the distance from the feature increases. For example, in a material, the screening environment (*i.e.* the importance of many-body effects) has a strong influence over how quickly this attenuation occurs. In metals it occurs quickly, but in large band-gap insulators the falloff is much more gradual. Even though quantum mechanics involves fully non-local operators, it has been noted that matter is, in practice, near-sighted. [54]

This idea of operator locality is the primary motivation for the subdivision technique used in EDNNs: an $L \times L$ training example is divided into $N = L^2/f^2$ non-overlapping regions of size $f \times f$. We call these regions focus. Then to each focus region, we provide overlapping context of width $c$. Each of these $N$ tiles (Fig. 2b) of size $(f + 2c) \times (f + 2c)$ is then fed into an identical neural network (Fig. 2a), and the $N$ individual outputs are summed to impose the extensivity of the operator. The loss is computed with respect to this final, summed value, and backpropagation is used to update the weights. In a normal domain-decomposition technique, some method to compensate for the inherent double-counting of the overlapping context regions would be necessary, however with EDNNs, we leave the task of rectifying this double-counting to the deep neural network itself; it must somehow learn to partially ignore the overlapping context regions.

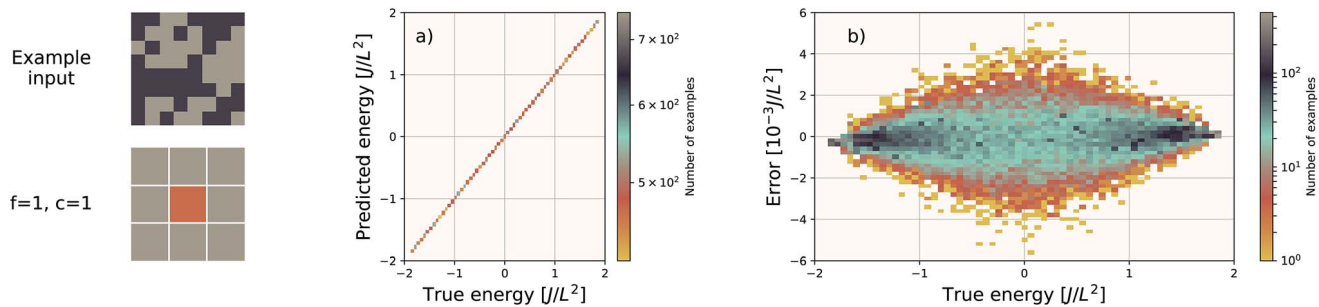

**Fig. 3** Performance of an EDNN tasked with learning the energy $E$ operator for the 8 × 8 Ising model. Since $E$ is semi-local ($l = 1$), $f = c = 1$ is an optimal configuration. Additional information in the form of a larger context region does not help the network predict values, and in fact makes the training more difficult, as the network must learn to ignore a significant amount of information. (a) Predicted *vs.* true energies (per spin) for optimal model. (b) Error (predicted − true energy) *vs.* true energy for optimal EDNN model.

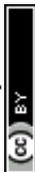

## 2.1 EDNN input, topology, and training

Our explanation of EDNNs was very general, with no information about the neural network we used, the loss function employed, or the representation of the input data used. This is intentional, as the crux of the EDNN technique is the way in which the input data is subdivided, and the individual contributions summed prior to backpropagation. In practice, creating an EDNN requires only slight (albeit fundamental) additions to the neural network topologies commonly used. In fact, there is no strict requirement that any neural network be used within the EDNN framework at all; any supervised machine learning model can be used. Furthermore, the technique can be applied to any representation of a spatially multidimensional field, provided spatial correlations are captured and consistent across the input representation, and the quantity of interest is extensive. This means that quantities such as entropy, magnetic moment, and electron density (in addition to energy as demonstrated in this work) are ideal candidates for the EDNN technique.

As for the "brain" of the model, we chose to demonstrate the technique with a neural network; the neural network within the EDNN can be of arbitrary complexity. We have made use of both deep convolutional neural networks as well as fully-connected artificial neural networks, depending on the complexity of the underlying problem. Since the input to the neural network is much smaller than the overall example, network architectures that would normally be prohibitively expensive become tractable with EDNNs (*e.g.* within the EDNN framework it would be possible to use a fully-connected artificial neural network to process a many-megapixel input).

In our case, we have focused on neural networks. Details of training hyperparameters are provided in Table 1. The fully-connected artificial neural network used in training the Ising model is comprised of a fully-connected layer of size $32(f + 2c)^2$ (note that $(f + 2c)$ is the size of a single tile), followed by two fully-connected layers with 64 outputs, which finally feed into a fully-connected layer of size 1. The convolutional deep neural network used to train the DFT models is constructed from 13 convolutional layers and two fully-connected layers. The first 2 layers are reducing and operate with filter sizes of $3 \times 3$ pixels. Each of these reducing layers operates with 64 filters and a stride of $2 \times 2$. The next 6 layers are non-reducing, meaning they have unit stride and preserve the resolution of the image. Each of these non-reducing layers operates with 16 filters of size $4 \times 4$. The ninth convolutional layer operates with 64 filters of size of $3 \times 3$ and a stride of $2 \times 2$. The last four convolutional layers are non-reducing, and operate with 32 filters of size $3 \times 3$. The final convolutional layer is fed into a fully-connected layer of size 1024. This layer feeds into a final fully-connected layer with a single output. The contribution from each tile is summed, and the loss is computed as the mean-squared error between this value and the true value.

We note that parallel branching within neural networks is a common technique, usually taking one of two forms. The first technique, which is used by *e.g.* GoogLeNet (a.k.a Inception),[55] uses repeating modules of parallel convolutional layers. Ref. 44 uses a similar approach, with multiple preprocessing techniques feeding a variety of data representations through many branches of a neural network architecture. Each branch of these neural networks has its own set of weights, and learns different features. Ultimately the output from each branch is concatenated to produce an ensemble of learned features that is subsequently fed into a decision layer.

The second approach employs a single set of weights, shared across the parallel branches such as in ref. 56. This technique facilitates efficient parallelization of the neural network training as each branch can be evaluated on separate hardware with little communication between devices. When training in parallel, the gradients from each separate branch are averaged, and the weights of each branch are updated synchronously. This effectively leads to a multiplicative speed-up in training and inference.

EDNNs are fundamentally different from these approaches, however, since in contrast to the former, the contribution from each branch is summed and not concatenated, and in contrast to the latter, the gradients used to update the weights are computed after the extensivity-imposing summation.

## 2.2 Focus and context

EDNNs introduce two new hyperparameters to the design of the neural network: the focus and the context sizes. The following considerations can ease in the choice of an appropriate focus and context:

• $c \approx l$ in order to provide sufficient context to the network, but should not be too much greater so as to introduce

**Table 1** Parameters used to train the various example models. Mean Squared Error (MSE) is used as the loss function in all examples

| Model | Total training examples | Batch size | Network | Optimizer | Learning rate | Loss | Epochs |
|---|---|---|---|---|---|---|---|
| Ising model | 100 000 | 2000 | DNN | Adam[57] | 0.0001 | MSE | 500 |
| Hexagonal sheets | 18 515 | 100 | CNN | Adam | 0.00001 | MSE | 500 |
| Porous graphene | 501 473 | 64 | CNN | Adam | $1 \times 10^{-6}$ | MSE | 500 |

 

redundant computations into the network. Context too large is not only inefficient for evaluation, but also makes the weight-optimization process more challenging as the network must learn to ignore a larger fraction of the input signal.

• Choosing $f$ is a balancing act between parallelizability and overall computational cost. Minimizing $f$ results in more tiles, which can be computed independently and thus parallelized efficiently. On the other hand, small focus leads to a greater overall computational demand, as for every focus region, there are overlapping context pixels that are being computed multiple times.

• A quantitative comparison of different focus and context pairs is difficult, as varying these parameters consequently changes the architecture of the neural network (e.g. number of weights, or required number of layers to reduce the image to a predetermined size through strided convolutions), modifying the fitting capabilities of the network.

We note that as the locality length scale of an operator grows, the optimal EDNN approaches a single tile being processed by a normal deep neural network; the context region is simply periodic padding. This is because for fully non-local operators, it is physically impossible to divide the problem up in the style of EDNNs since the full volume is needed to make an inference. Such systems are rare in practice, thankfully. EDNNs therefore represent a framework that can naturally handle the full continuum of possible screening environments. EDNNs can describe all phases of the electron gas. We note that when dealing with operators that have large values of $l$, it is often useful to recast the problem in reciprocal space, and such an approach could be useful too with EDNNs.

## 3 Results

As illustrative examples of the method, we have trained an EDNN on three systems: (1) the ferromagnetic Ising model and (2) quantum mechanical total energy calculations (within the density functional theory framework) for (a) hexagonal systems (e.g. boron nitride, graphite, and heterostructures of the two), and (b) porous graphene sheets.

### 3.1 Example: the Ising model

The Ising model is a two-state spin ($\sigma$) model with $\sigma = \pm 1$ with a total energy Hamiltonian

$$\hat{H} = -J \sum_{\langle i,j \rangle} \sigma_i \sigma_j \tag{1}$$

where $J$ is the interaction strength, and the summation is computed over nearest-neighbour pairs ($\langle i,j \rangle$). For $J = 1$ the system is ferromagnetic; it is favourable for neighbouring spins to align. Application of EDNN to the Ising model is particularly instructive because the locality length scale of the Hamiltonian operator is known explicitly; as previously discussed, it is an $l = 1$ operator. The nearest-neighbour interaction means that $c > 1$ provides no additional information to the EDNN. Including this data makes the task more difficult, as the network must learn to completely ignore these features. As the size of the context region grows beyond the locality length scale of the operator, the learning process is less efficient. Our optimal EDNN trained on the Ising model, using raw, binary spin values ($\sigma \sim \{1, -1\}$) as input, achieves a MAE (median absolute error) of $0.028J/L^2$ on the testing set, sufficiently accurate[40] to reproduce the finite temperature phase transition for which the model is so well-known. In comparison, the Ising models subdivided using unit focus with $c = 2$ and $c = 3$ produce an error of $1.090J/L^2$ and $14.816J/L^2$, respectively. All three EDNNs were trained for the same number of iterations and since the layer size of the artificial neural network is dependent upon the input size, larger context means significantly more parameters in the neural network. With $f = 1$, $c = 3$ the neural network contains over 7 times the parameters as the $f = 1$, $c = 1$ neural network and therefore is much more difficult to optimize. This is another motivation for considering carefully an appropriate choice for $f$ and $c$.

### 3.2 Example: density functional theory

#### 3.2.1 Hexagonal sheets. For comparison with previous work, we reuse a previously reported dataset[42] of 2d crystalline structures. This dataset consists of 26 449 structures of crystalline and defect (missing atom) hexagonal surfaces. The technique of EDNNs does not depend on the atomic representation used as input to the neural network; provided the spatial structure is properly represented; we use the previously established atomic representation using an image generated from the function

$$V(x, y) = \sum_{i=1}^{N} Z_i \exp\left(-\left[\frac{(x - x_i)^2 + (y - y_i)^2}{2\gamma^2}\right]\right). \tag{2}$$

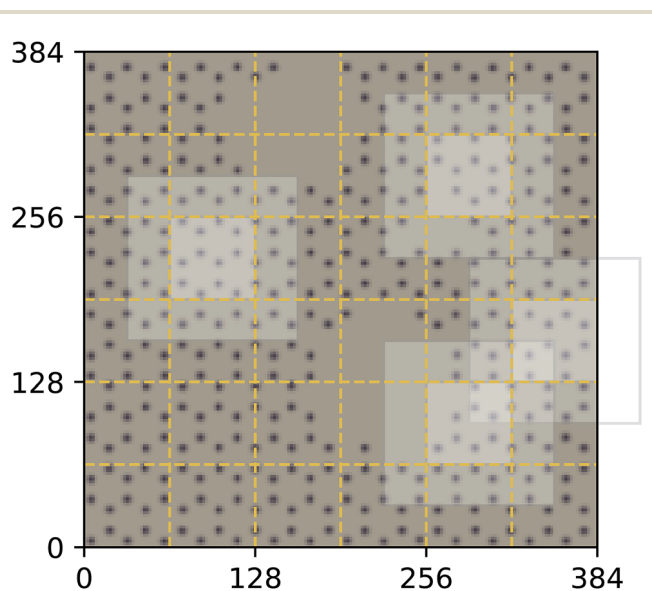

**Fig. 4** Decomposition of input image for the quantum mechanical density functional theory calculation using $f = 64$ and $c = 32$. Four tiles consisting of a focus region and context region are highlighted. Overlap in the context region is by design and the EDNN must learn to ignore this overlap in the final reduction of the extensive quantity.

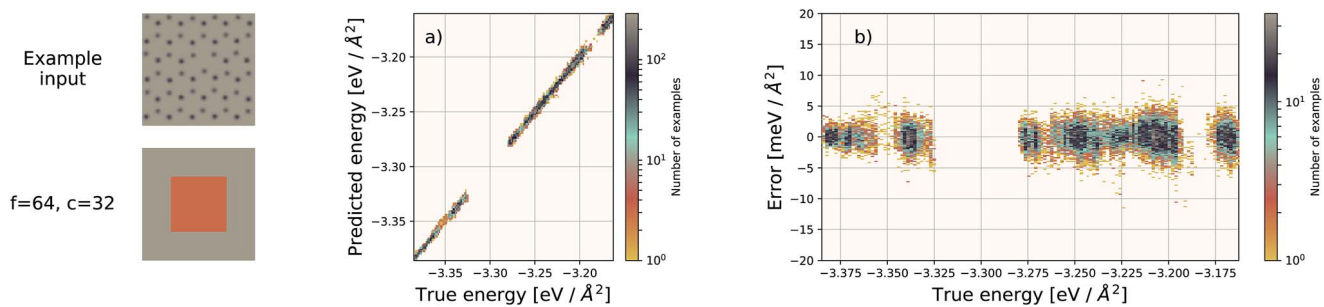

Fig. 5 Left: an example graphene sheet. (a) Performance of our EDNN model on a testing set (predicted vs. true). (b) Error ((predicted − true) vs. true) for the EDNN model trained on the total energy as calculated through the density functional theory framework.

Here $x_i$, $y_i$ are the coordinates of the $i^{th}$ atom with atomic number $Z_i$. $\gamma = 0.2$ Å was used as the width of the atomic potential wells, consistent with Brockherde *et al.*[39] (Fig. 4). This function is evaluated on a $256 \times 256$ grid representing a 12.5 Å $\times$ 13 Å unit cell containing 60 atoms arranged in a hexagonal lattice. The dataset contains both graphene and hexagonal boron nitride, with and without defects.

For these quantum mechanical systems (computed within the generalized gradient approximation of the density functional theory[48,49] framework), it is not possible to determine a specific value of $l$. Within the hierarchy of approximations, the method we used to compute the total energy (PBE[51]) is considered to be a semi-local approximation, since the exchange–correlation potential includes only gradients of the charge density. Other terms within the total energy are fully non-local (although they are subject to screening by the electron gas). Nonetheless, total energy is an extensive quantity and again the EDNN performs favourably compared to previously reported (non-EDNN) results.

Using $f = 64$ and $c = 32$, we achieve a MAE of 1.122 meV Å$^{-2}$ on our test set after 500 training epochs, notably better than the MAE of 2.529 meV Å$^{-2}$ on a traditional (non-EDNN) network (Fig. 5).[42] For this choice of ($f$, $c$) our input representation is divided into 16 tiles, enabling an inference speed-up factor of 16 due to the ability to calculate the contribution from each tile in parallel. This is on top of the inherent speed-up of evaluating the EDNN compared to DFT. A conservative estimate for this latter speed-up is on the order of 1 million in terms of CPU-hours.

The ability of an EDNN to learn to ignore, or compensate for redundant context can be explored by measuring the performance of the model when information is partially obfuscated through the application of a Gaussian blur to select regions of the input. In Fig. 6, we plot the performance of the network when blur is applied within the context region (edge) during inference. As expected, when the blur is applied at the periphery of the context region, the network reports very similar values for $\varepsilon$ as when there is no blur present. As the blurring encroaches on more context, the predictions become poorer; the neural network is evidently learning to ignore the context region. This is to be expected, as the data will appear again in the focus region of another tile. This is how double-counting is avoided. When a small area in the center of the focus region is blurred, the neural network is able to make accurate predictions, (likely due to the limited amount of information being lost), but as the extent of the blur increases within the focus region, the accuracy of inference suffers greatly.

### 3.2.2 Porous graphene sheets.

As a more challenging example of the applicability of this technique, we developed a dataset of porous graphene sheets. We generated 3137 starting geometries by randomly removing varying numbers of regions of various sizes from pristine graphene sheets of size 35 Å $\times$ 35 Å. We separated the starting configurations into one set of 349 starting configurations, reserved for testing the ability of the trained EDNN to generalize to data that it has never seen (validation), and the remaining 2788 configurations were used for generating a training set.

We ran molecular dynamics at a temperature of 1000 K using forces obtained through the density functional theory framework (using VASP[58-61] and the PBE[51] functional), collecting configurations from the molecular dynamics trajectories as training. In all, we collected 501 473 training configurations and 60 744 testing configurations.

We use the same Gaussian-based input representation, and the same deep convolutional neural network architecture as the "hexagonal sheets" investigation. The larger supercell size of the porous graphene sheets requires discretization on a larger grid; we chose a $384 \times 384$ grid. This results in more tiles required, but the training procedure remains identical in all other aspects. The results are shown in Fig. 7. The EDNN performs well with a median absolute error of 1.685 meV Å$^{-2}$.

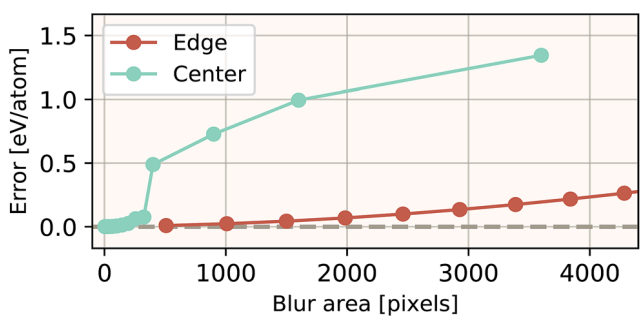

Fig. 6 Resilience of an EDNN to the addition of obfuscating Gaussian blur. When a constant amount of information (constant area) within the tiles is blurred, examples that had context area blurred result in more accurate inference than examples that had focus blurred. This is evidence that the EDNN is learning to ignore the context regions.

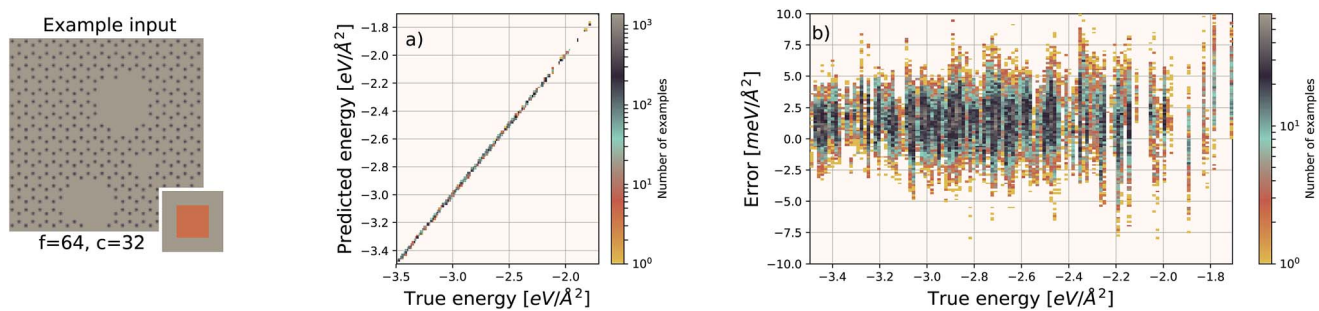

**Fig. 7** Left: an example porous graphene sheet. (a) The true (DFT) vs. predicted (EDNN) total energy in eV Å$^{-2}$. The tight clustering along the diagonal indicates the EDNN performs well at predicting the total energy. (b) The error (DFT energy – EDNN energy), in meV Å$^{-2}$, is very close to zero.

### 3.3 Transferability to arbitrary input size

EDNNs have the capability of making predictions on input configurations of arbitrary size so long as the physical length scale (i.e. the real space extent of a pixel in the input representation), is preserved, and the input size remains an integer multiple of the focus region.

To demonstrate this, we used the neural network trained on $8 \times 8$ Ising configurations to make energy predictions of $128 \times 128$ Ising configurations sampled near the critical temperature. Without any further training, the median absolute error on these much larger configurations was $2.055 J/L^2$. This is substantially larger than the $8 \times 8$ error, but this is to be expected. Since there is some error associated with the prediction of a given tile, it indeed makes sense that this error will scale with the extensivity of the system. In other words, the error relative to the absolute energy is still small. The predicted values and the prediction error are plotted against the true values in Fig. 8.

Additionally we test the DFT model trained on a 12.5 Å × 13 Å, $256 \times 256$ grid ($N = 60$ atoms) on several larger domains up to 62.6 Å × 65.1 Å ($1024 \times 1024$ grid, $N = 1500$ atoms). During inference, the EDNN performs similarly well predicting energies of configurations larger than those in the training set, as seen in Fig. 9, and does so many orders of magnitude faster than conventional numerical methods. This is a powerful feature of EDNN, as one can generate a testing set of many training examples for significantly less computational expense, and then apply it to larger systems without the $\mathcal{O}(N^3)$ (or worse) scaling inherent in Kohn–Sham density functional theory. The evaluation of the EDNN scales as $\mathcal{O}(N)$. The fact that an EDNN can take a $\mathcal{O}(N^3)$ problem and map it to $\mathcal{O}(N)$ might seem suspicious at first. Recall though that HK DFT does scale linearly and that the polynomial scaling of Kohn–Sham DFT is due to the diagonalization of the Kohn–Sham Hamiltonian. We avoid such an evaluation during inference, and therefore we achieve scaling consistent with orbital free DFT.

As a proof of concept and to further demonstrate the exceptional scaling of the EDNN approach, we generated a porous graphene sheet comprised of 35.2 million atoms, with a supercell size of 1.0 μm². EDNN inference is trivially parallel, so using a custom distributed TensorFlow implementation, we were able to compute the total energy of the sheet using 448 cores across 16 nodes in 24.7 minutes. A "ground truth" DFT calculation at this scale is intractable, but based on the results on smaller-scale tests (Fig. 8 and 9) we can confidently conclude that the relative error is comparable to that of a DFT method. These results are shown in Fig. 10 and 11.

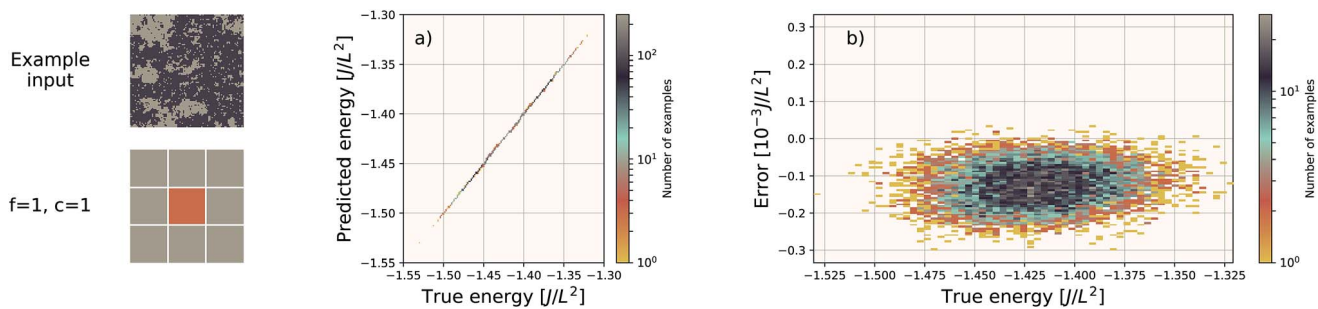

**Fig. 8** An EDNN trained only on $8 \times 8$ Ising training examples is capable of making accurate predictions of the $128 \times 128$ Ising model near criticality. While the absolute error is higher at $2.055 J/L^2$, the relative error is very small. While it appears the EDNN consistently overpredicts the energy, this is not an effect of large scale inference, but rather that the input configurations are from an energy window where the original EDNN also slightly overpredicted the energy. This is evident when compared to the appropriate region of Fig. 3b.

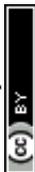

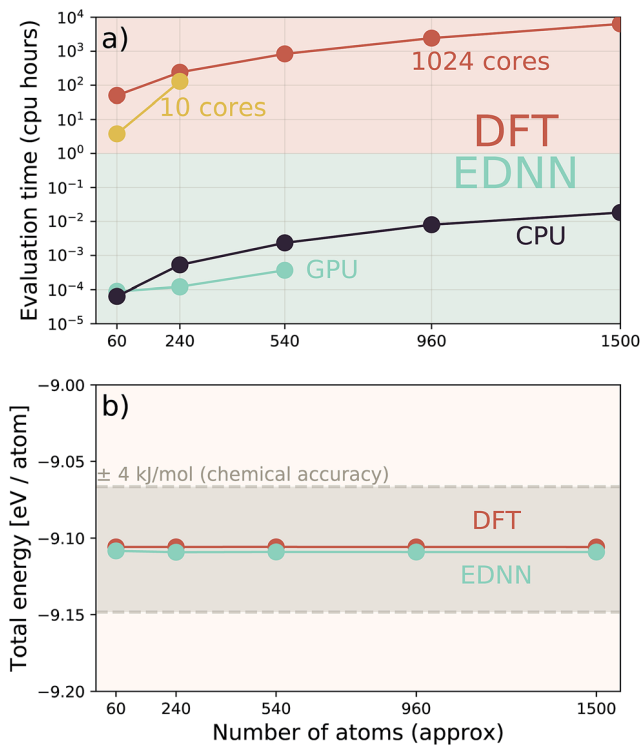

**Fig. 9** A single EDNN was trained on a 12.5 Å × 13 Å unit cell. This trained model was used to make accurate predictions on larger unit cells not present in the training set. (a) The inference time for large systems was about 1 million times smaller than the equivalent density functional theory approach, with CPU evaluation performing better than GPU evaluation on large systems. (b) The resulting energy predictions are consistent within chemical accuracy of 1 kcal mol⁻¹. The scale of the error can be expected to scale linearly with the size of the system (*i.e.* $\mathcal{O}(N)$).

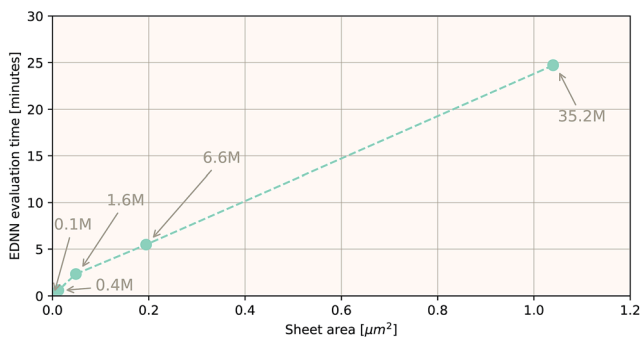

**Fig. 10** Using the model trained on many small porous graphene sheets, we used a multi-node, distributed TensorFlow implementation to make predictions on large sheets. The model evaluation time scales linearly with the cell area (and thus, under the assumption of homogeneous density, the number of atoms). The annotations refer to the number of atoms in the configuration. The EDNN allows for total energy calculations at DFT accuracy of more than 1.0 μm² of material in 24.7 minutes (Fig. 12). Importantly, the model was trained on a dataset of configurations consisting of only around 500 atoms, and therefore collection of training data does not require accurate simulation of large configurations. All EDNN evaluations were carried out on a 20-node cluster with 28 cores per node.

### 3.4 EDNN in use

We have demonstrated that EDNNs can be used to accurately learn the mapping from atomic coordinates to energy. How though, do EDNNs perform when used as the energy evaluation function in an actual simulation? To investigate this we performed a Metropolis Monte Carlo (MC) simulation of pristine graphene, the same 12.5 Å × 13 Å unit cell used in the dataset from Section 3.2.1 above. During the MC calculation we use the EDNN to evaluate the energy of the atomic configuration. The accurate evaluation of energy is important within the MC framework, as the evolution of the atom positions depends exponentially on the accurate evaluation of energy. In Fig. 13, we plot the radial distribution function, $g(r)$ for the atoms in the EDNN (MC) simulation alongside $g(r)$ for a molecular dynamics (MD) simulation of pristine graphene using DFT (VASP). For the MD, we used the same VASP input parameters as the MD used to generate the EDNN training/testing dataset. Both simulations occur at a temperature of 1000 K. The two radial distribution functions are in close agreement, with the peaks in exact agreement. While there is a bit of discrepancy between the two functions, it is evident that the EDNN technique can be used to perform physically relevant simulations at a fraction of the cost of the quantum mechanical alternative; during the MC simulation the EDNN takes 0.39 seconds per energy-evaluation, while the DFT (MD) takes 9.9 seconds. The bulk of the EDNN calculation time is spent within our inefficient image representation construction, *i.e.* the evaluation of eqn (2). Since this is not the focus of this work, we have not yet optimized this evaluation, so considerably higher performance is possible in practice.

### 3.5 EDNN advantages

We note that EDNNs are particularly well suited for massive parallelization, particularly during inference, since neural network evaluation can be distributed across hardware; there is no need for communication between them until the final summation. Beyond scalability, EDNNs have the feature that they can operate on inputs of arbitrary shape and size (to within integer multiples of *f*). This is particularly useful for treating large-scale mesoscopic structures *in silico*.

Since the neural network of the EDNN only operates on a single tile at a time, EDNNs permit the use of more computationally intensive network architectures (*e.g.* fully-connected networks), that would normally be infeasible. Additionally EDNNs are well suited for Monte–Carlo sampling, as local updates would only require the re-evaluation of nearby tiles, not the entire configuration.

Under the EDNN framework, there is no requirement to assign energy to a particular region of space or atomic species, unlike atom-centred methods. Rather, the network is simply told that the extensive property applies to the entire system. This is important because it is extremely flexible; it allows for a seamless method that can learn quantum molecular mechanics, implicit solvation energies, entropy, *etc.* Furthermore EDNNs can operate on any spatial field, such as the electron density.

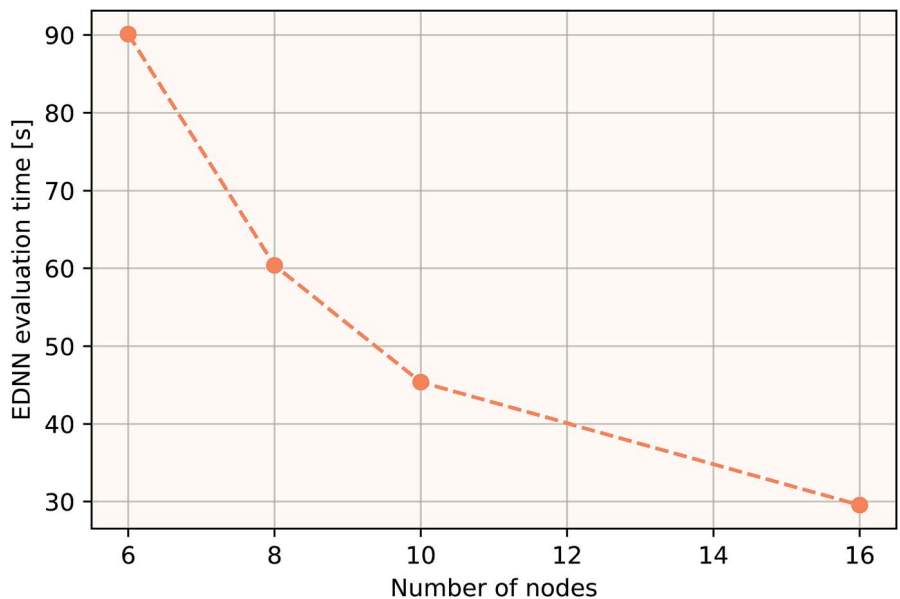

**Fig. 11** Using the model trained on many small porous graphene sheets, we used a multi-node implementation of TensorFlow to perform inference on larger systems. At over 400 000 atoms, we achieve better-than-linear scaling, even with only typical gigabit ethernet interconnect. In theory, since the evaluation of an EDNN is perfectly subdivisible into separate parts, with the only communication cost incurred during the final summation, scaling to large system sizes should be parallel. In practice, overhead is incurred in the distribution of input data, but we achieve impressive scaling nonetheless.

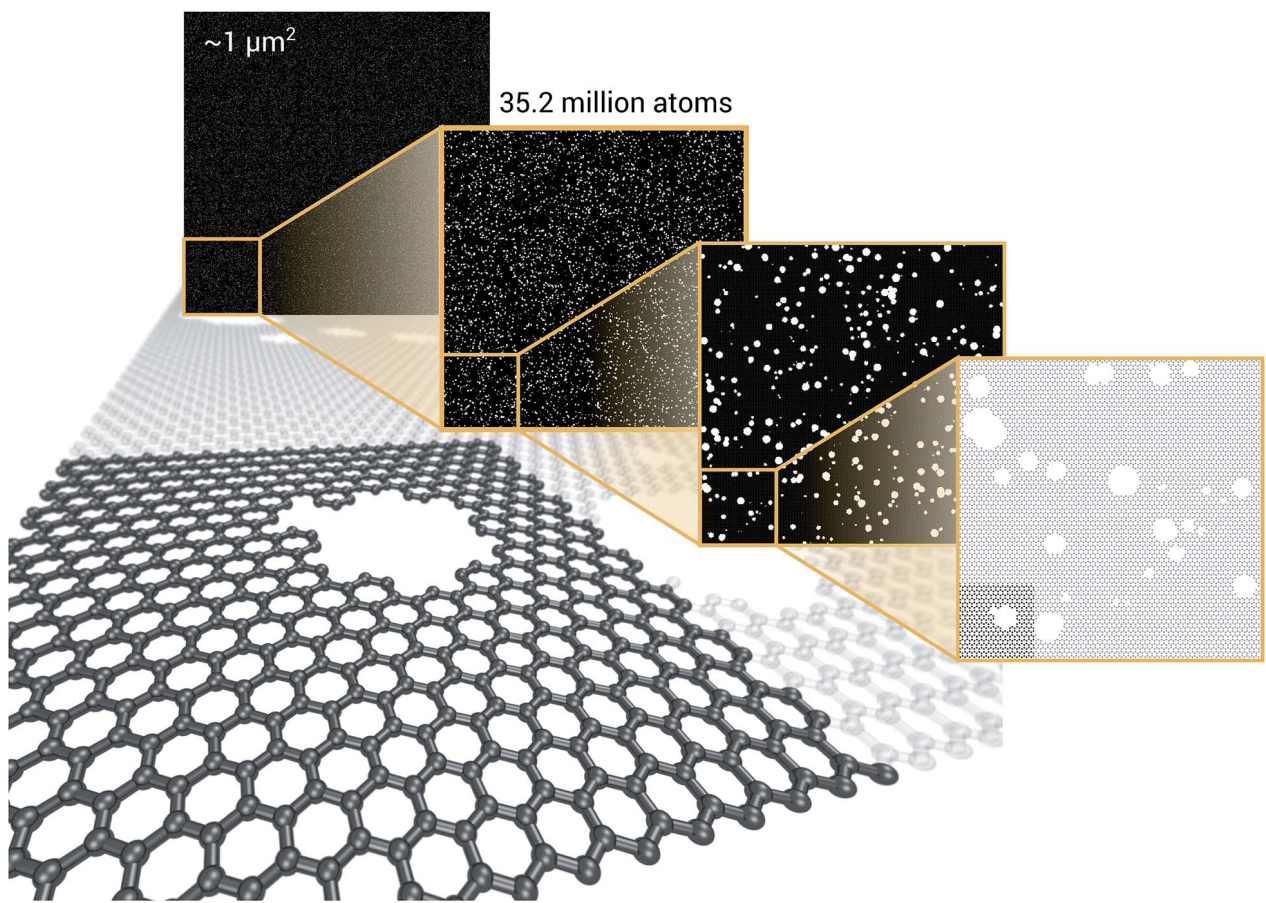

**Fig. 12** We demonstrate that EDNNs can be used to make an energy prediction of a two-dimensional 35.2 million atom system, over 1.0 μm² of material, at an accuracy comparable to density functional theory, in under 25 minutes. Additionally, the evaluation of the neural network scales linearly with the number of atoms (assuming relatively homogeneous density), so this evaluation-time estimate can be driven lower with wider hardware configurations. Such a system exists on a length scale visible with optical microscopy and larger than some living organisms.

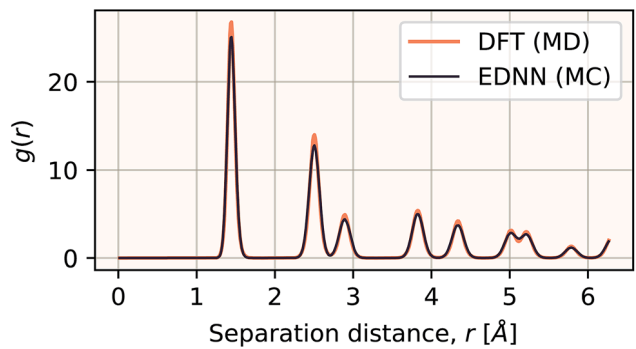

**Fig. 13** The radial distribution functions $g(r)$ plotted for two calculations. The black line is $g(r)$ for a Metropolis Monte Carlo simulation using the EDNN as the energy evaluation function. The orange line is $g(r)$ for a molecular dynamics calculation using density functional theory as the energy evaluation criteria. Since the two methods differ algorithmically, a direct comparison is difficult, but we can see that both methods yield exactly the same peak positions, indicating the EDNN is capable of making predictions at an accuracy suitable for performing physical simulations.

EDNN can be thought of as a generalization of a complex, many-body force-field. Features are learned automatically by the EDNN, and domain decomposition is handled intrinsically. Traditional force-fields assume that extensive properties such as the total energy can be expressed as a many-body sum over interacting particles, and in some cases, an implicit solvation environment. While many different models exist, almost all share the common feature of expanding the total energy in terms of neighbour interactions, typically within a fixed cutoff radius. Bond lengths, angles, and partial charges are used as features, and the coefficients of the relative terms are trained against a fixed set of examples. Even methods designed for metallic environments (*e.g.* the embedded atom method) make use of structural "features" (*e.g.* the density of neighbours) which are then fed into a feature based decision algorithm. When used for atomistic modelling, EDNN accomplish the same task as a force-field without the requirement of hand selecting features. They are also extremely straightforward to implement in parallel, and do not require complex calculations of angles, dihedrals, feature vectors, or neighbour lists for efficient parallelization.

### 3.6 EDNNs capturing physics

At the most fundamental level, many-body interactions within a material are subject to screening. Screening occurs due to elementary excitations that occur within the system (*e.g.* electronic, rotational, *etc.*). Depending on the characteristic of the electron gas, screening effects can result in a rapid interaction decay length (*e.g.* as in a metal), or they may attenuate much more slowly. Screening and the various length scales at which its effects are observed are emergent phenomena.

EDNNs are built on the idea that interactions are screened at some length scale, but that *a priori*, the user does not know what it is. The training data itself encodes this length scale, and the network takes advantage of it. The generality of the concept and the implementation are why EDNNs are so useful; the physical property that permits the decomposition of the problem is actually revealed by the data itself.

On a similar note, (*i.e.* that the data should reveal the physics, rather than incorporating it *a priori*) relates to the question of invariance; there are currently two schools of thought about how symmetries should be built into models. Within the chemical literature, there is currently a strong bias toward enforcing symmetries within models first, and then developing methods that are constrained by those symmetries. This "symmetry first" view is generally not what has been the approach in the computer vision/artificial intelligence community. We are of the belief that symmetries (*e.g.* rotations) can be learned and should not necessarily be enforced. This is both from a pragmatic point of view (too many constraints on a network during the learning process can restrict it to local minima), and from somewhat of a philosophical point of view. In deep learning, features can and should be learned from the dataset itself. The way to teach a neural network a physical law is to provide it data from which it can learn.

## 4 Conclusion

We have demonstrated a new form of deep neural network, motivated by physics, that can operate on arbitrary sized input and physical length scales while maintaining the extensivity of properties inferred by the network. Networks of this form are particularly well suited to large-scale parallel inference, as the individual components of the input data can be computed independently of one another. We demonstrate the ability of EDNNs to learn extensive operators, such as the nearest-neighbour Ising Hamiltonian and the density functional theory total energy operator. The process of optimizing the focus and context hyperparameters provides physical insight into the interaction length scales of the physical problem. We demonstrate the efficiency of the EDNN approach in inferring properties of systems much larger than those on which it was trained. Finally, we demonstrate the ability of an EDNN to infer the total energy of a porous graphene sheet comprised of 35.2 million atoms, to DFT accuracy, in under 25 minutes. Although we have chosen to demonstrate three specific examples in the field of physics, the techniques and arguments that we present are quite general, and naturally apply to many problems in physics and image processing.

## Conflicts of interest

There are no conflicts to declare.

## Data and code

The porous graphene data set, as well as links to EDNN tutorials and code can be found at http://doi.org/10.4224/c8sc04578j.data.

## Appendix

Let us formalize the EDNN technique.

The (two dimensional, single channel) EDNN takes as input a batch of images $x \in \mathbb{R}^{N_B \times L \times L \times 1}$, where $N_B$ is the batch size. EDNNs are not limited to 2-dimensional single-channel data and can be used for three (or more) dimensions and multiple channels.

We construct the input tensor $\mathbf{X}$ by taking a periodically-padded copy of $\mathbf{x}$ and performing a strided slice, starting from the origin with stride equal to the focus. We extract a patch of size $(f + 2c) \times (f + 2c)$, e.g.

$$\mathbf{X}_{b,i,j,d} = \mathbf{x}_{b,if-c:if+f+c,jf-c:jf+f+c,d}, \ i,j \in [1\ldots L/f]$$

meaning

$$\mathbf{X} \in \mathbb{R}^{N_B \times (L/f) \times (L/f) \times d \times (f+2c) \times (f+2c)}$$

In words, $\mathbf{X}$ is comprised of $L^2/f^2$ tiles of size $(f + 2c)^2$ $d$-channel pixels for each of the $N_B$ images in a batch.

Each tile is passed individually through the approximation function (e.g. a neural network), which is parametrized by $\theta$:

$$\mathcal{C}_{b,i,j} = f\left(X_{b,i,j}; \theta\right).$$

We call the $\mathcal{C}$ tensor the "tile contributions", that is, how much of the final answer is contained in each tile. The tile contributions are reduced through a summation, preserving only the batch dimension:

$$\hat{\varepsilon}_b = \sum_i^{L/f} \sum_j^{L/f} \mathcal{C}_{b,i,j}.$$

This vector, $\hat{\varepsilon}_b$, is the predicted extensive quantity from the EDNN, one entry for each example in the input batch.

The batch loss is computed as the mean-squared error between the predicted and the true value (i.e. the "label"):

$$\mathcal{L} = \frac{1}{N_B} \sum_{i=0}^{N_B} \left(\hat{\varepsilon}_i - \varepsilon_i\right)^2$$

This loss function is then minimized as in a normal neural network, using some form of gradient descent to tweak the parameters $\theta$ so that the prediction matches the true answer as closely as possible.

## Acknowledgements

The authors acknowledge funding from NSERC and SOSCIP. Compute resources were provided by Compute Canada, SOSCIP, National Research Council of Canada, and an NVIDIA Faculty Hardware Grant.

## Notes and references

1 S. Chetlur and C. Woolley, arXiv, 2014, **1–9**.

2 G. Lacey, G. W. Taylor and S. Areibi, arXiv, 2016.

3 Y. Jia, E. Shelhamer, J. Donahue, S. Karayev, J. Long, R. Girshick, S. Guadarrama and T. Darrell, arXiv, 2014, 675–678.

4 D. Silver, A. Huang, C. J. Maddison, A. Guez, L. Sifre, G. van den Driessche, J. Schrittwieser, I. Antonoglou, V. Panneershelvam, M. Lanctot, S. Dieleman, D. Grewe, J. Nham, N. Kalchbrenner, I. Sutskever, T. Lillicrap, M. Leach, K. Kavukcuoglu, T. Graepel and D. Hassabis, *Nature*, 2016, **529**, 484–489.

5 V. Mnih, K. Kavukcuoglu, D. Silver, A. a. Rusu, J. Veness, M. G. Bellemare, A. Graves, M. Riedmiller, A. K. Fidjeland, G. Ostrovski, S. Petersen, C. Beattie, A. Sadik, I. Antonoglou, H. King, D. Kumaran, D. Wierstra, S. Legg and D. Hassabis, *Nature*, 2015, **518**, 529–533.

6 G. Carleo and M. Troyer, *Science*, 2017, **355**, 602–606.

7 X. Liang, W.-Y. Liu, P.-Z. Lin, G.-C. Guo, Y.-S. Zhang and L. He, *Phys. Rev. B: Condens. Matter Mater. Phys.*, 2018, **98**, 104426.

8 O. S. Ovchinnikov, S. Jesse, P. Bintacchit, S. Trolier-Mckinstry and S. V. Kalinin, *Phys. Rev. Lett.*, 2009, **103**, 2–5.

9 L. F. Arsenault, A. Lopez-Bezanilla, O. A. von Lilienfeld and A. J. Millis, *Phys. Rev. B: Condens. Matter Mater. Phys.*, 2014, **90**, 155136.

10 K. Ch'ng, J. Carrasquilla, R. G. Melko and E. Khatami, *Phys. Rev. X*, 2017, **7**, 031038.

11 E. P. L. van Nieuwenburg, Y.-H. Liu and S. D. Huber, *Nat. Phys.*, 2017, **13**, 435–439.

12 J. Carrasquilla and R. G. Melko, *Nat. Phys.*, 2017, **13**, 431–434.

13 R. B. Jadrich, B. A. Lindquist and T. M. Truskett, *J. Chem. Phys.*, 2018, **149**, 194109.

14 X. L. Zhao and L. B. Fu, arXiv, 2018.

15 D. Kim and D.-h. Kim, *Phys. Rev. E*, 2018, **98**, 022138.

16 T. Bereau, R. A. DiStasio, A. Tkatchenko and O. A. von Lilienfeld, *J. Chem. Phys.*, 2018, **148**, 241706.

17 A. G. Kusne, T. Gao, A. Mehta, L. Ke, M. C. Nguyen, K. Ho, V. Antropov, C.-Z. Wang, M. J. Kramer, C. Long and I. Takeuchi, *Sci. Rep.*, 2015, **4**, 6367.

18 S. Jesse, M. Chi, A. Belianinov, C. Beekman, S. V. Kalinin, A. Y. Borisevich and A. R. Lupini, *Sci. Rep.*, 2016, **6**, 26348.

19 P. V. Balachandran, D. Xue, J. Theiler, J. Hogden and T. Lookman, *Sci. Rep.*, 2016, **6**, 19660.

20 N. Artrith and A. Urban, *Comput. Mater. Sci.*, 2016, **114**, 135–150.

21 J. N. Wei, D. Duvenaud and A. Aspuru-Guzik, *ACS Cent. Sci.*, 2016, **2**, 725–732.

22 R. Gómez-Bombarelli, J. N. Wei, D. Duvenaud, J. M. Hernández-Lobato, B. Sánchez-Lengeling, D. Sheberla, J. Aguilera-Iparraguirre, T. D. Hirzel, R. P. Adams and A. Aspuru-Guzik, *ACS Cent. Sci.*, 2018, **4**, 268–276.

23 L. Messina, A. Quaglino, A. Goryaeva, M.-c. Marinica, C. Domain, N. Castin, G. Bonny and R. Krause, arXiv, 2018.

24 K. T. Schütt, H. E. Sauceda, P.-J. Kindermans, A. Tkatchenko and K.-R. Müller, *J. Chem. Phys.*, 2018, **148**, 241722.

25 J. Behler and M. Parrinello, *Phys. Rev. Lett.*, 2007, **98**, 146401.

26 M. Hellström and J. Behler, *Phys. Chem. Chem. Phys.*, 2017, **19**, 82–96.

27 J. S. Smith, O. Isayev and A. E. Roitberg, *Chem. Sci.*, 2017, **8**, 3192–3203.

28 V. L. Deringer and G. Csányi, *Phys. Rev. B: Condens. Matter Mater. Phys.*, 2017, **95**, 094203.

29 K. Yao and J. Parkhill, *J. Chem. Theory Comput.*, 2016, **12**, 1139–1147.

30 A. Ziletti, D. Kumar, M. Scheffler and L. M. Ghiringhelli, *Nat. Commun.*, 2018, **9**, 1–10.

31 E. J. Kim and R. J. Brunner, *Mon. Not. R. Astron. Soc.*, 2016, **000**, 1–13.

32 A. Aurisano, A. Radovic, D. Rocco, A. Himmel, M. Messier, E. Niner, G. Pawloski, F. Psihas, A. Sousa and P. Vahle, *J. Instrum.*, 2016, **11**, P09001.

33 R. Acciarri, C. Adams, R. An, J. Asaadi, M. Auger, L. Bagby, B. Baller, G. Barr, M. Bass, F. Bay, M. Bishai, A. Blake, T. Bolton, L. Bugel, L. Camilleri, D. Caratelli, B. Carls, R. C. Fernandez, F. Cavanna, H. Chen, E. Church, D. Cianci, G. Collin, J. Conrad, M. Convery, J. Crespo-Anadón, M. Del Tutto, D. Devitt, S. Dytman, B. Eberly, A. Ereditato, L. E. Sanchez, J. Esquivel, B. Fleming, W. Foreman, A. Furmanski, G. Garvey, V. Genty, D. Goeldi, S. Gollapinni, N. Graf, E. Gramellini, H. Greenlee, R. Grosso, R. Guenette, A. Hackenburg, P. Hamilton, O. Hen, J. Hewes, C. Hill, J. Ho, G. Horton-Smith, C. James, J. J. de Vries, C.-M. Jen, L. Jiang, R. Johnson, B. Jones, J. Joshi, H. Jostlein, D. Kaleko, G. Karagiorgi, W. Ketchum, B. Kirby, M. Kirby, T. Kobilarcik, I. Kreslo, A. Laube, Y. Li, A. Lister, B. Littlejohn, S. Lockwitz, D. Lorca, W. Louis, M. Luethi, B. Lundberg, X. Luo, A. Marchionni, C. Mariani, J. Marshall, D. M. Caicedo, V. Meddage, T. Miceli, G. Mills, J. Moon, M. Mooney, C. Moore, J. Mousseau, R. Murrells, D. Naples, P. Nienaber, J. Nowak, O. Palamara, V. Paolone, V. Papavassiliou, S. Pate, Z. Pavlovic, D. Porzio, G. Pulliam, X. Qian, J. Raaf, A. Rafique, L. Rochester, C. R. von Rohr, B. Russell, D. Schmitz, A. Schukraft, W. Seligman, M. Shaevitz, J. Sinclair, E. Snider, M. Soderberg, S. Söldner-Rembold, S. Soleti, P. Spentzouris, J. Spitz, J. St. John, T. Strauss, A. Szelc, N. Tagg, K. Terao, M. Thomson, M. Toups, Y.-T. Tsai, S. Tufanli, T. Usher, R. Van de Water, B. Viren, M. Weber, J. Weston, D. Wickremasinghe, S. Wolbers, T. Wongjirad, K. Woodruff, T. Yang, G. Zeller, J. Zennamo and C. Zhang, *J. Instrum.*, 2017, **12**, P03011.

34 W. Bhimji, S. A. Farrell, T. Kurth, M. Paganini, Prabhat and E. Racah, *J. Phys.: Conf. Ser.*, 2018, **1085**, 042034.

35 A. P. Bartók, R. Kondor and G. Csányi, *Phys. Rev. B: Condens. Matter Mater. Phys.*, 2013, **87**, 184115.

36 Y. Levine, O. Sharir, N. Cohen and A. Shashua, *Phys. Rev. Lett.*, 2019, **122**, 065301.

37 T. Xie and J. C. Grossman, *Phys. Rev. Lett.*, 2018, **120**, 145301.

38 K. Yao, J. E. Herr, D. W. Toth, R. Mckintyre and J. Parkhill, *Chem. Sci.*, 2018, **9**, 2261–2269.

39 F. Brockherde, L. Vogt, L. Li, M. E. Tuckerman, K. Burke and K.-R. Müller, *Nat. Commun.*, 2017, **8**, 872.

40 K. Mills and I. Tamblyn, *Phys. Rev. E*, 2018, **97**, 032119.

41 K. Mills, M. Spanner and I. Tamblyn, *Phys. Rev. A*, 2017, **96**, 042113.

42 K. Ryczko, K. Mills, I. Luchak, C. Homenick and I. Tamblyn, *Comput. Mater. Sci.*, 2018, **149**, 1–18.

43 N. Portman and I. Tamblyn, *J. Comput. Phys.*, 2017, **43**.

44 D. Ciresan, U. Meier and J. J. Schmidhuber, Computer Vision and Pattern Recognition (CVPR) *IEEE Conference on 2012*, 2012, pp. 3642–3649.

45 A. K. Rappe, C. J. Casewit and K. S. Colwell, *J. Am. Chem. Soc.*, 1992, **2**, 10024–10035.

46 K. T. Schütt, F. Arbabzadah, S. Chmiela, K. R. Müller and A. Tkatchenko, *Nat. Commun.*, 2017, **8**, 13890.

47 W. Pronobis, K. T. Schütt, A. Tkatchenko and K.-R. Müller, *Eur. Phys. J. B*, 2018, **91**, 178.

48 P. Hohenberg and W. Kohn, *Phys. Rev.*, 1964, **136**, B864–B871.

49 W. Kohn and L. J. Sham, *Phys. Rev.*, 1965, **140**, A1133–A1138.

50 J. Perdew, J. Chevary, S. Vosko, K. Jackson, M. Pederson, D. Singh and C. Fiolhais, *Phys. Rev. B: Condens. Matter Mater. Phys.*, 1992, **46**, 6671–6687.

51 J. P. Perdew, M. Ernzerhof and K. Burke, *J. Chem. Phys.*, 1996, **105**, 9982–9985.

52 C. Lee, W. Yang and R. G. Parr, *Phys. Rev. B: Condens. Matter Mater. Phys.*, 1988, **37**, 785–789.

53 A. D. Becke, *J. Chem. Phys.*, 1993, **98**, 5648–5652.

54 E. Prodan, *Phys. Rev. B: Condens. Matter Mater. Phys.*, 2006, **73**, 085108.

55 C. Szegedy, W. Liu, Y. Jia, P. Sermanet, S. Reed, D. Anguelov, D. Erhan, V. Vanhoucke and A. Rabinovich, *Proceedings of the IEEE Computer Society Conference on Computer Vision and Pattern Recognition*, 2015, pp. 1–9.

56 K. Alex, I. Sutskever and G. E. Hinton, *Neural Information Processing Systems (NIPS)*, 2012, pp. 1097–1105.

57 D. P. Kingma and J. Ba, arXiv, 2014, **1–15**.

58 G. Kresse and J. Hafner, *Phys. Rev. B: Condens. Matter Mater. Phys.*, 1993, **47**, 558–561.

59 G. Kresse and J. Hafner, *Phys. Rev. B: Condens. Matter Mater. Phys.*, 1994, **49**, 14251–14269.

60 G. Kresse and J. Furthmüller, *Comput. Mater. Sci.*, 1996, **6**, 15–50.

61 G. Kresse, *Phys. Rev. B: Condens. Matter Mater. Phys.*, 1996, **54**, 11169–11186.